\documentclass[namedreferences]{solarphysics}

\usepackage[hyperref,optionalrh]{spr-sola-addons} 
\usepackage[all]{hypcap} 
\usepackage{graphicx}        
\usepackage{amssymb}        
\usepackage{textcmds}        
\usepackage{color}           
\usepackage{breakurl}        

\usepackage{psfrag}
\usepackage{auto-pst-pdf}


\usepackage{setspace}
\hypersetup{
	colorlinks = true, 
	allcolors = blue,
}

\makeatletter
\newcommand\tsb[1]{\@textsubscript{\selectfont#1}}
\def\@textsubscript#1{{\m@th\ensuremath{_{\mbox{\fontsize\sf@size\z@#1}}}}}
\newcommand\tsp[1]{\@textsuperscript{\selectfont#1}}
\def\@textsuperscript#1{{\m@th\ensuremath{^{\mbox{\fontsize\sf@size\z@#1}}}}}

\providecommand{\e}[1]{\ensuremath{\times 10^{#1}}}
\providecommand{\lsr}[1]{\textit{V}\tsb{LSR}}

\providecommand{\rsolar}[1]{#1 R\tsb{$\odot$}}

\usepackage{multirow}

 \renewcommand*{\thetabnote}{\alph{tabnote}}

\providecommand{\edit}[1]{{\color{black}{#1}}}

\begin{document}


\psfrag{X (rsun)}[][][0.8]{{\fontfamily{phv}\selectfont X (R$_{\odot}$)}}
\psfrag{Y (rsun)}[][][0.8]{{\fontfamily{phv}\selectfont Y (R$_{\odot}$)}}
\psfrag{Height (rsun)}[][][0.8]{{\fontfamily{phv}\selectfont Height (R$_{\odot}$)}}

\begin{article}

\begin{opening}

\title{Densities Probed by Coronal Type III Radio Burst Imaging}

\author[addressref={af1},corref,email={patrick.mccauley@sydney.edu.au}]{\inits{P.~I.}\fnm{Patrick~I.}~\lnm{McCauley}\orcid{0000-0002-1450-7350}}
\author[addressref={af1},corref,email={iver.cairns@sydney.edu.au}]{\inits{I.~H.}\fnm{Iver~H.}~\lnm{Cairns}\orcid{0000-0001-6978-9765}}
\author[addressref={af2},corref,email={john.morgan@curtin.edu.au}]{\inits{J.}\fnm{John}~\lnm{Morgan}}

\address[id=af1]{School of Physics, University of Sydney, Sydney, NSW 2006, Australia}
\address[id=af2]{International Centre for Radio Astronomy Research, Curtin University, Perth, WA 6845, Australia}

\runningauthor{McCauley, Cairns, \& Morgan}
\runningtitle{Densities Probed by Coronal Type III Radio Burst Imaging}


\begin{abstract}

We present coronal density profiles derived from low-frequency (80--240 MHz) imaging of 
three type III solar radio bursts observed at the limb by the Murchison Widefield Array (MWA). 
Each event is associated with a white light streamer at larger heights and is plausibly associated 
with thin extreme ultraviolet rays at lower heights. 
Assuming harmonic plasma emission, we find average electron densities 
of 1.8\e{8} cm\tsp{-3} down to 0.20\e{8} cm\tsp{-3} at heights of 1.3 to \rsolar{1.9}. 
These values represent $\sim$2.4--5.4$\times$ enhancements over canonical background levels and 
are comparable to the highest streamer densities obtained from data at other wavelengths. 
Assuming fundamental emission instead would increase the densities by a factor of 4. 
High densities inferred from type III source heights can be explained by assuming that the exciting electron 
beams travel along overdense fibers or by radio propagation effects 
\edit{that may cause a source to appear at a larger height than the true emission site.}
We review the arguments for both scenarios in light of recent results.
We compare the extent of the quiescent corona to model predictions to estimate the 
impact of propagation effects, which we conclude can only partially explain the apparent density enhancements. 
Finally, we use the time- and frequency-varying source positions to estimate electron beam 
speeds of between 0.24 and 0.60 c. 

\end{abstract}
\end{opening}


\section{Introduction} %
\label{introduction} %

Type III solar radio bursts are caused by semi-relativistic electrons 
streaming through and perturbing the ambient coronal or interplanetary plasma. 
A recent review is given by \citet{Reid14}.
The dominant theory, proposed by \citet{Ginzburg58}, invokes a two-step 
process beginning with the stimulation of Langmuir waves (plasma oscillations) 
in the background plasma by an electron beam. 
A small fraction of the Langmuir wave energy is then converted into electromagnetic 
radiation at either the local electron plasma frequency ($f_p$) or its harmonic 
(2$f_p$; see reviews by \citealt{Robinson00,Melrose09}).
The emission frequency depends mainly on the ambient electron density ($n_e$) because 
$f_p \propto \sqrt{n_e}$. 
This relationship produces the defining feature of type III bursts, a rapid drift from high to low frequencies 
as the exciter beam travels away from the Sun through decreasing densities \citep{Wild50}. 

The rate at which the emission frequency drifts ($df/dt$) is therefore related to the 
electron beam speed, which can be obtained in the radial direction by assuming a density model $n_e(r)$. 
Many authors have employed this technique for various events with various models, generally 
finding modest fractions of light speed (0.1--0.4 $c$; e.g. \citealt{Alvarez73,Aschwanden95,Mann99,Melendez99,Krupar15,Kishore17}).  
Alternatively, the coronal and/or interplanetary density gradient can be inferred by instead assuming a  
beam speed (e.g. \citealt{Fainberg71,Leblanc98}) or by simply assuming that the beam speed is constant \citep{Cairns09}. 
While these methods can yield robust estimates for the density gradient, they cannot be converted into an explicit 
density structure $n_e(r)$ without normalizing the gradient to a specific value at a specific heliocentric distance. 
This normalization has typically been done using estimates from white light polarized brightness data close to the Sun, 
\textit{in situ} data in the interplanetary medium, or the observed height of type III burst sources at various 
frequencies. 

Densities inferred from type III source heights, particularly at lower frequencies, 
have frequently conflicted with those obtained from other methods. 
The earliest spatial measurements found larger source heights than would be expected from 
fundamental plasma emission, implying density enhancements of an order of magnitude or more \citep{Wild59}. 
This finding was confirmed by subsequent investigations (e.g. \citealt{Morimoto64,Malitson66}), and along 
with other arguments, led many authors to 
two conclusions: 
First, that harmonic (2$f_p$) emission likely dominates
(e.g. \citealt{Fainberg71,Mercier74,Stewart76}). 
This brings the corresponding densities down by a factor of 4, then implying only a moderate enhancement 
over densities inferred from white light data. 
(Counterarguments for the prevalence of fundamental emission will be referenced in Section~\ref{density}.) 
Second, that the electron beams preferentially traverse overdense flux tubes
(e.g. \citealt{Bougeret84}), a conclusion bolstered by 
spatial correlations between several type III bursts and white light streamers 
(e.g. \citealt{Trottet82,Kundu84,Gopalswamy87,Mugundhan18}). 

The overdense hypothesis has been challenged by evidence that the large 
source heights can instead be explained by propagation effects.
If type III emission is produced in thin, high-density structures, then it can escape relatively 
unperturbed through its comparatively rarefied surroundings.  
However, if the emission is produced in an environment near the associated plasma level 
(i.e. with an average $n_e$ corresponding to the radio waves' equivalent $f_p$), then refraction and scattering 
by density inhomogeneities may substantially shift an observed source from its 
true origin (e.g. \citealt{Leblanc73,Riddle74,Bougeret77}). 
\citet{Duncan79} introduced the term \textit{ducting} in this context, 
\edit{which refers to emission being guided to larger heights within a low-density structure though 
successive reflections against the high-density ``walls" of the duct.} 
This concept was generalized for a more realistic corona by \citet{Robinson83}, who showed that random 
scattering of radio waves by thin, overdense fibers has the \edit{same} net effect of elevating an observed source 
radially above its emission site.
\edit{Additional details on this topic, along with coronal refraction, will be given in Section~\ref{propagation}.}

Many authors came to favor propagation effects instead of the overdense structure interpretation 
for a few reasons. 
Despite the aforementioned case studies, type IIIs did not appear to be statistically 
associated with regions of high average density in the corona \citep{Leblanc74,Leblanc77} or 
in the solar wind \citep{Steinberg84}. 
Interplanetary (kHz-range) type III source regions are also so large as to demand angular broadening 
by propagation effects (e.g. \citealt{Steinberg85,Lecacheux89}). 
Invoking propagation effects can also 
be used to explain apparent spatial differences between fundamental and harmonic sources (e.g. \citealt{Stewart72,Kontar17})\edit{, 
along with large offsets between radio sources on the disk and their likely electron acceleration sites (e.g. \citealt{Bisoi18}). 
These arguments are reviewed by \citet{Dulk00}, and further discussion with additional recent references will be presented 
in Sections~\ref{propagation} and \ref{discussion}.} 

\edit{Both the interpretation of electron beams moving along overdense structures and of radio propagation effects 
elevating burst sources} rely on the presence of thin, high-density fibers. 
Either the electron beams are traveling within these structures or the type III emission is being scattered 
by them. In this paper, we will suggest that propagation effects are important but cannot entirely 
explain the density enhancements for our events.
Section~\ref{observations} describes our observations: \ref{mwa} outlines our data reduction, 
\ref{events} details our event selection criteria, and \ref{context} describes the multi-wavelength 
context for the selected type III bursts. 
Section~\ref{analysis} describes our analysis and results: \ref{density} infers densities from type III 
source heights, \ref{speed} estimates electron beam speeds from imaging data, and 
\ref{propagation} examines propagation effects by comparing the extent of the quiescent 
corona to model predictions. 
In Section~\ref{discussion}, we discuss the implications of our results, along with other recent 
developments, on the debate between the overdense and propagation effects hypotheses. 
Finally, our conclusions are summarized in Section~\ref{conclusion}. 

 
\section{Observations} %
\label{observations} %


\subsection{Murchison Widefield Array (MWA)}
\label{mwa}

The MWA is a low-frequency radio interferometer in Western Australia 
with an instantaneous bandwidth of 30.72 MHz that can be flexibly distributed 
from 80 to 300 MHz \citep{Tingay13}. 
Our data were recorded with a 0.5 s time cadence and a 40 kHz spectral 
resolution, which we average over 12 separate 2.56 MHz bandwidths 
centered at 
80, 89, 98, 108, 120, 132, 145, 161, 179, 196, 217, and 240 MHz.
We use the same data processing scheme as \citet{McCauley17}, and 
what follows is a brief summary thereof. 

Visibilities were generated with the standard MWA correlator \citep{Ord15} and the  
\texttt{cotter} software \citep{Offringa12,Offringa15}. 
Observations of bright and well-modelled calibrator sources were used 
to obtain solutions for the complex antenna gains \citep{Hurley14}, which 
were improved by imaging the calibrator and iteratively self-calibrating from there \citep{Hurley17}. 
\texttt{WSClean} \citep{Offringa14} was used to perform the imaging 
with a Briggs -2 weighting \citep{Briggs95} to maximize spatial resolution and minimize point spread function (PSF) sidelobes. 
The primary beam model of \citet{Sutinjo15} was used to produce Stokes I 
images from the instrumental polarizations, 
and the SolarSoftWare (SSW\footnote{SSW: \url{https://www.lmsal.com/solarsoft/}}, \citealt{Freeland98}) 
routine \texttt{mwa\_prep} \citep{McCauley17} was used to translate the images onto solar coordinates. 
Flux calibration was achieved by comparison with thermal bremsstrahlung and gyroresonance emission predictions from 
FORWARD\footnote{FORWARD: \url{https://www2.hao.ucar.edu/modeling/FORWARD-home}} \citep{Gibson16} 
based on the Magnetohydrodynamic Algorithm outside a Sphere model
(MAS\footnote{MAS: \url{http://www.predsci.com/hmi/data\_access.php}}; \citealt{Lionello09}). 


 \begin{figure}
 \centerline{\includegraphics[width=\textwidth,clip=]{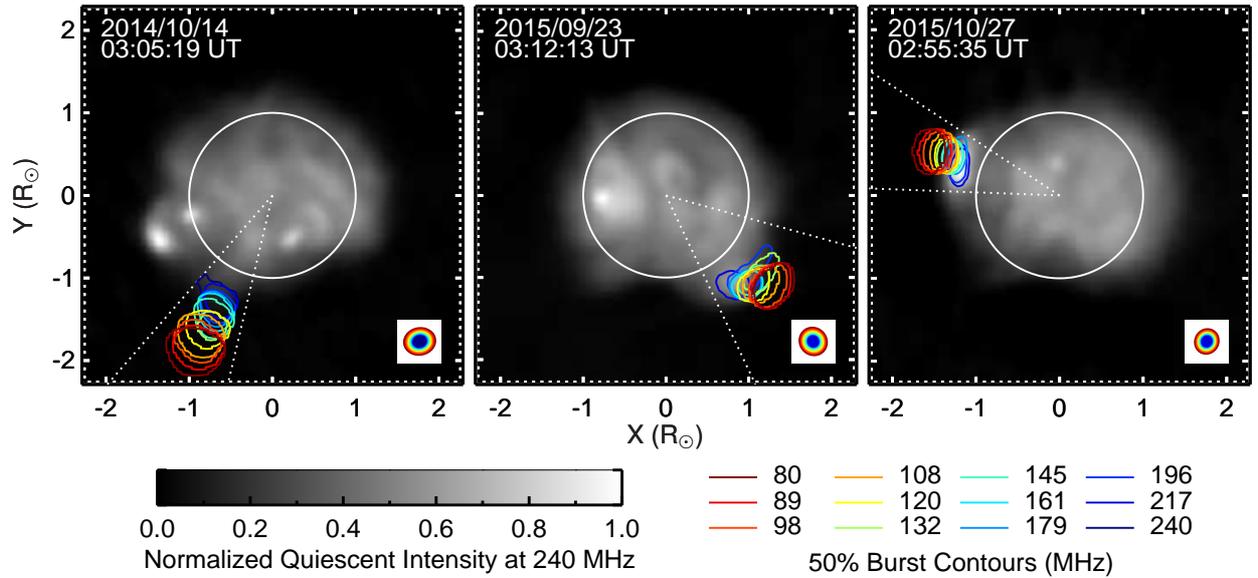}}
 \caption{MWA type III burst contours at 50\% of the peak intensity 
 for each channel overlaid on 240 MHz images of the quiescent corona. 
 The solid circle represents the optical disk, and dotted lines bound the region included 
 in the dynamic spectra (Figs.~\ref{fig:ds_comp} \& \ref{fig:spectra}). 
 Colored ellipses in the lower-right corners show the synthesized beam sizes for each channel.}
 \label{fig:centroids}
 \end{figure}


 \begin{figure}
 \centerline{\includegraphics[width=\textwidth,clip=]{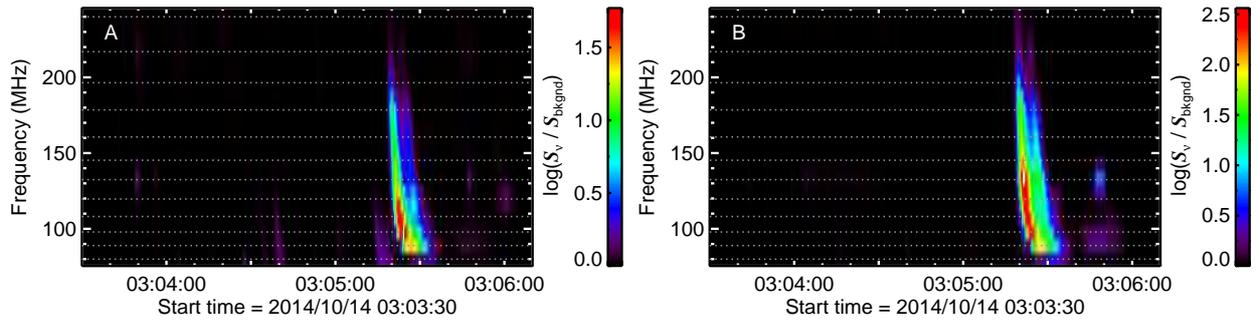}}
 \caption{Dynamic spectra constructed from image intensities 
 for the type III burst near 03:05:20 UT on 2014-10-14. 
 Panel A includes the full FOV, while B includes only 
 the segment bounded by the dotted lines in Fig.~\ref{fig:centroids}.
Dotted horizontal lines show the locations of the 12 channels, each having a spectral width of 2.56 MHz. 
Intensities have been divided by the background level and plotted a logarithmic scale. 
 }
 \label{fig:ds_comp}
 \end{figure}


\begin{figure}
\centerline{\includegraphics[width=\textwidth,clip=]{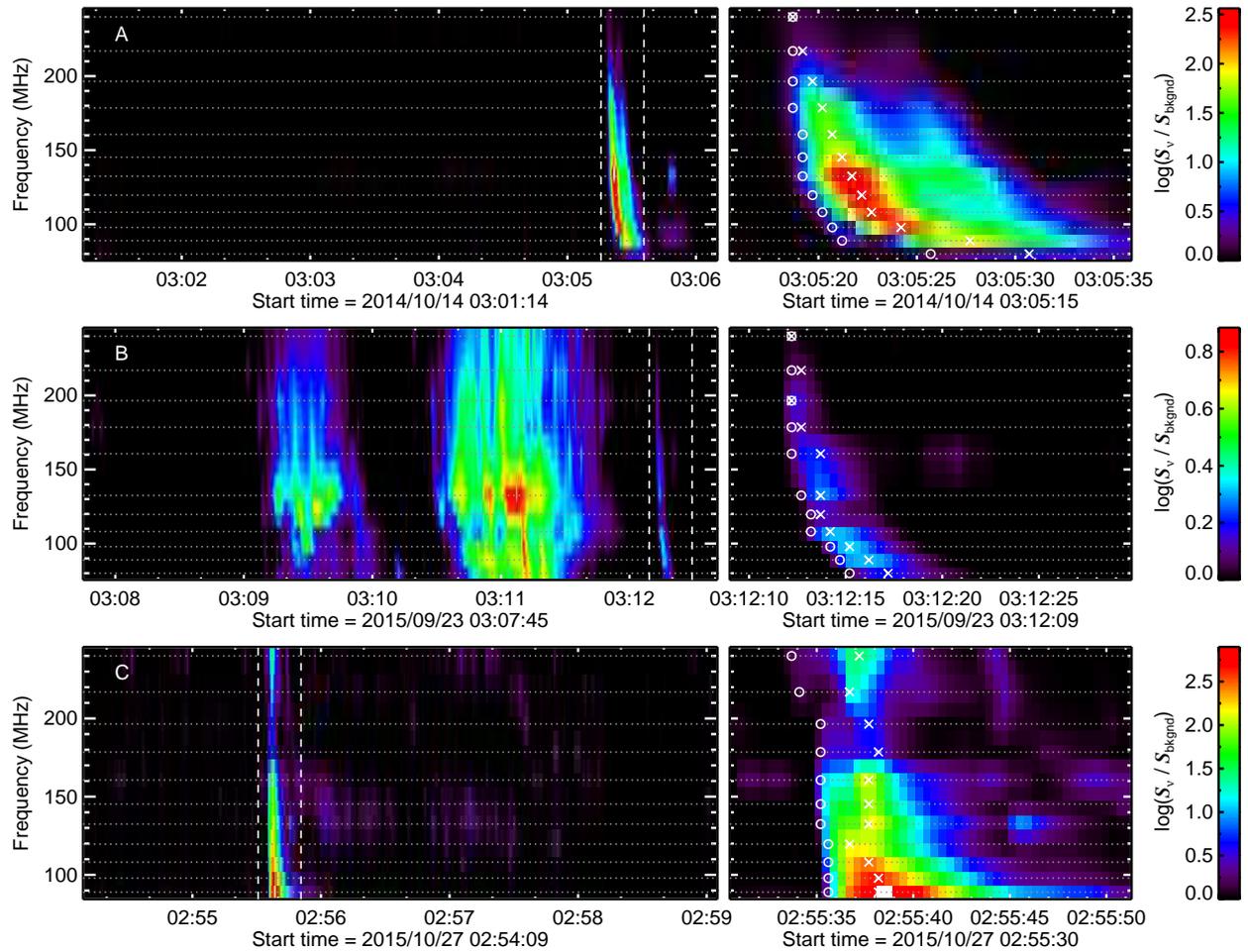}}
\caption{Dynamic spectra constructed from partial image intensities, including only the FOV segment bounded 
by the dotted lines in Fig.~\ref{fig:centroids}. 
The left column shows the full 5-min observation intervals, while the right column shows 20-sec 
periods surrounding the selected type III bursts. 
Circles and crosses denote the onset and peak burst times for each channel.  
}
\label{fig:spectra}
\end{figure}


\subsection{Event Selection}
\label{events}

These data are part of an imaging survey of many
type III bursts observed by the MWA during 45 separate observing 
periods in 2014 and 2015. 
\citet{McCauley17} performed a case study of an event that exhibits 
unusual source motion, and future work will present statistical analyses. 
Burst periods during MWA observing runs were identified using the daily 
National Oceanic and Atmospheric Administration (NOAA) solar event 
reports\footnote{NOAA event reports: \url{http://www.swpc.noaa.gov/products/solar-and-geophysical-event-reports}}
based on observations from the \textit{Learmonth}  \citep{Guidice81,Kennewell03}
and \textit{Culgoora} \citep{Prestage94} solar radio spectrographs, which overlap with 
the MWA's frequency range at the low and high ends, respectively. 

Three events were selected from the full sample based on the following criteria. 
First, the burst sites needed to be located at the radio limb with roughly 
radial progressions across frequency channels. 
Limb events minimize projection effects, allowing us to reasonably approximate the 
projected distance from Sun-center as the actual radial height. 
Second, to eliminate potential confusion between multiple events and to maximize spectral coverage,
the bursts needed to be sufficiently isolated in time and frequency, with a 
coherent drift from high to low frequencies across the full MWA bandwidth. 
Third, the source regions needed to be relatively uncomplicated ellipses with 
little-to-no intrinsic motion of the sort described by \citet{McCauley17}. 
This again minimizes projection effects and ensures that we follow a single 
beam trajectory for each event.  

Figure~\ref{fig:centroids} shows the burst contours for each channel overlaid on 
quiescent background images at 240 MHz.
Each of the three events occurred on a different day, and we refer to them by the 
UTC date on which they occurred. 
Figure~\ref{fig:ds_comp} shows dynamic spectra for the 2014-10-14 event, with the left panel  
covering the full Sun and the right panel including only the region demarcated by the dotted 
lines in Figure~\ref{fig:centroids}. 
The partial Sun spectrum excludes a neighboring region that is active over the 
same period, allowing the type III frequency structure to be more easily followed. 
This approach is similar to that of \citet{Mohan17}, who discuss the utility of spatially resolved dynamic spectra. 
Figure~\ref{fig:spectra} shows the masked spectra for all three events. 


\begin{figure}
\centerline{\includegraphics[width=\textwidth,clip=]{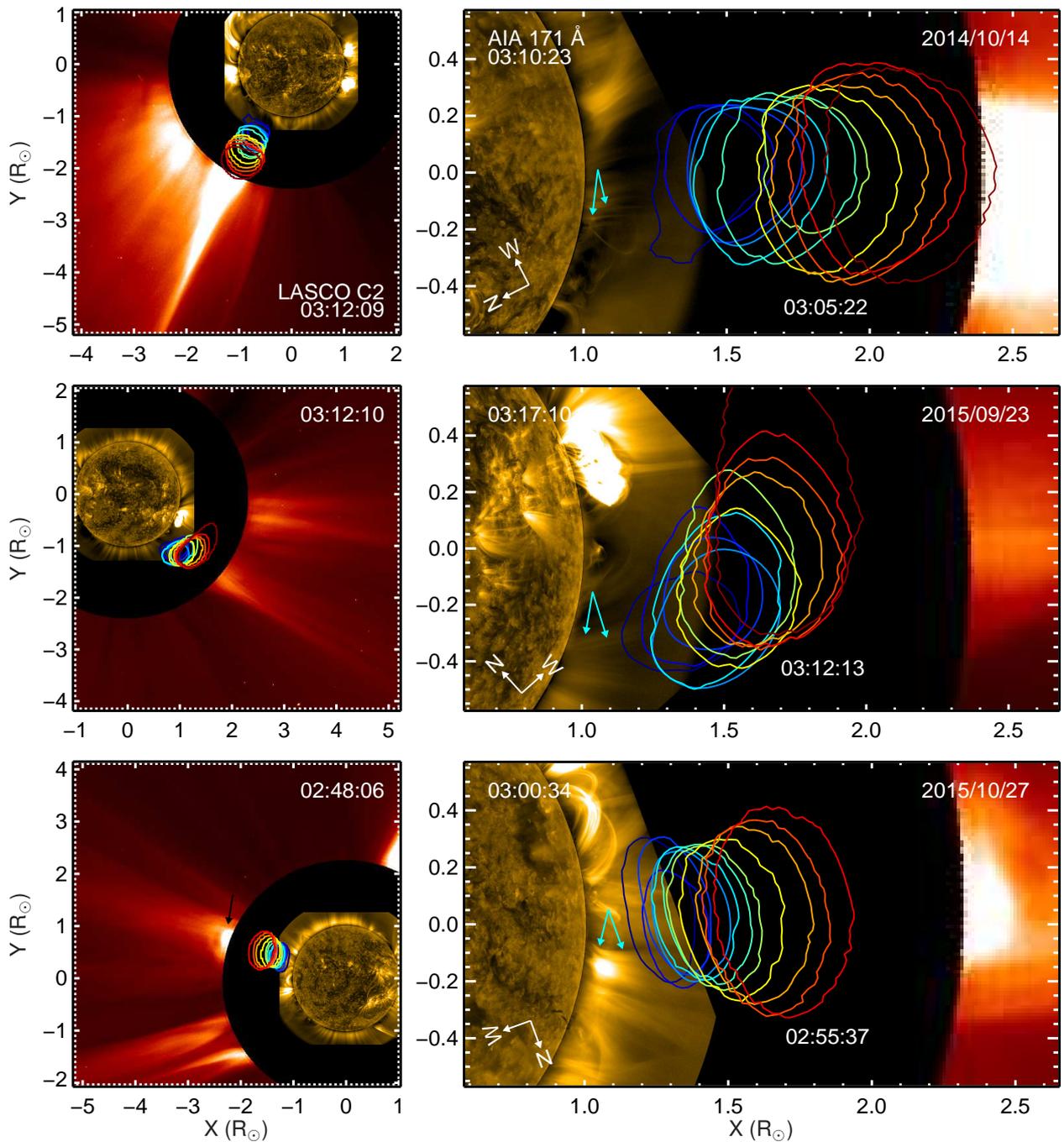}}
\caption{Overlays of the 50\% burst contours onto 
AIA 171 \AA{} and LASCO C2 images. 
Contour colors are for spectral channels from 80--240 MHz as in Fig.~\ref{fig:centroids}.
\edit{UTC observation times are shown for LASCO in the left panel, for AIA in the upper-left 
of the right panel, and for MWA in the middle of the right panel. 
The MWA times reflect the average peak time across frequency channels (see Fig.~\ref{fig:spectra}). 
The AIA images are 10-min (50-image) averages processed with a radial filter to accentuate off-limb features; 
times reflect the middle of these 10-min windows, which begin at the burst onsets and cover the subsequent periods 
over which associated EUV signatures would be expected.}
Images are rotated in the right column such that the burst progression 
is roughly horizontal, which helps illustrate the extent to which each event 
progresses radially. 
Cyan arrows point to the EUV structures that exhibit activity  
during or just after the radio bursts. 
\edit{The black arrow in the lower-left panel points to a CME that originated behind the limb and 
passed the C2 occulting disk around 20-min prior to the type III burst.}  
}
\label{fig:overlay}
\end{figure}

\subsection{Context}
\label{context}

In this section, we briefly describe the context for each of the radio bursts with 
respect to observations at other wavelengths and associated phenomena. 
Figure~\ref{fig:overlay} overlays the burst contours from Figure~\ref{fig:centroids} onto 
contemporaneous extreme ultraviolet (EUV) and white light data. 
The white light images were produced by the 
Large Angle and Spectrometric C2 Coronagraph (LASCO C2; \citealt{Brueckner95}) 
onboard the Solar and Heliospheric Observatory (SOHO; \citealt{Domingo95}).  
\edit{C2 has an observing cadence of 20 min, and Figure~\ref{fig:overlay} 
includes the nearest images in time to our radio bursts.}

The EUV data come from the 
Atmospheric Imaging Assembly (AIA; \citealt{Lemen12}) 
onboard the Solar Dynamics Observatory (SDO; \citealt{Pesnell12}).
We use the 171 \AA{} AIA channel, which is dominated by Fe IX 
emission produced by plasma at around 0.63 MK, because it most 
clearly delineates the fine magnetic structures along which type III beams 
are expected to travel. 
To further accentuate off-limb features, we apply a radial filter 
using the SSW routine \texttt{aia\_rfilter} \citep{Masson14}. 
Note that the apparent brightness of a given pixel in a radial filter image corresponds to  
its true intensity relative only to pixels of the same radial height (i.e. equally bright structures 
at different heights do not have the same physical intensity).
\edit{AIA has an observing cadence of 12 s, and Figure~\ref{fig:overlay} uses 10-min (50-image) 
averages that cover the periods during and immediately after the radio bursts. 
This time window is used because a potential EUV signature associated with a 
type III burst will propagate at a much lower speed than the burst-driving electron beam and 
will likely be most apparent in the minutes following the burst (e.g. \citealt{McCauley17,Cairns17}).}

In all cases, the radio bursts appear to be aligned with dense structures visible to AIA at lower 
heights and to LASCO C2 at larger heights. 
The latter case is obvious, with each set of burst contours situated just below bright 
white light streamers. 
Cyan arrows in the right panels of Figure~\ref{fig:overlay} identify the associated EUV structures, each 
of which exhibits a mild brightening and/or outflow during or immediately after the corresponding radio burst. 
This activity may be indicative of weak EUV jets, which are frequently associated 
with type III bursts (e.g. \citealt{Chen13,Innes16,McCauley17,Cairns17}), but robust outflows are not observed here. 
The alignment between the EUV and radio burst structure is particularly 
striking for the 2015-09-23 event in that both appear to follow roughly the same non-radial arc. 
A correspondence between EUV rays and type III bursts was previously reported by \citet{Pick09}. 
 
Type III bursts are commonly, but not always, associated with X-ray flares (e.g. \citealt{Benz05,Benz07,Cairns17}) and occasionally with 
Coronal Mass Ejections (CMEs; e.g. \citealt{Cane02,Cliver09}). 
Our 2014-10-14 event is not associated with either, but the other two are. 
On 2015-09-23, a weak B-class flare occurred just to the north of 
our radio sources from active region 12415.  
The flare peaked around 3:11 UT, which corresponds to a period of relatively 
intense coherent radio emission that precedes the weaker burst of interest here (see Figure~\ref{fig:spectra}). 
Given the radio source positions and associated EUV structure, we do not believe the flare site 
to be the source of accelerated electrons for our event, though the flare may have been responsible 
for stimulating further reconnection to the south. 

\edit{On 2015-10-27, a CME was ongoing at the time of the radio burst, and its leading edge, 
indicated by the black arrow in the lower left panel of Figure~\ref{fig:overlay}, 
can be seen just above the C2 occulting disk. 
Inspection of images from the Extreme Ultraviolet Imager (EUVI; \citealt{Howard08}) onboard the 
STEREO-A spacecraft shows that the CME originated from a large active region close to the east limb 
but occulted by the disk from AIA's perspective. 
The CME was launched well before our type III burst, but the region that produced it was very active 
over this period and is likely connected to the activity visible to AIA immediately after the radio burst 
along the structure indicated by the cyan arrows in the lower right panel of Figure~\ref{fig:overlay}. 
So while we do not think the CME was directly involved in triggering the radio burst, it may have impacted 
the medium through which the type III electron beam would later propagate, which is relevant 
to a hypothesis proposed by \citet{Morosan14} that will be discussed in Section~\ref{discussion}.}


\begin{figure}
\centerline{\includegraphics[width=\textwidth,clip=]{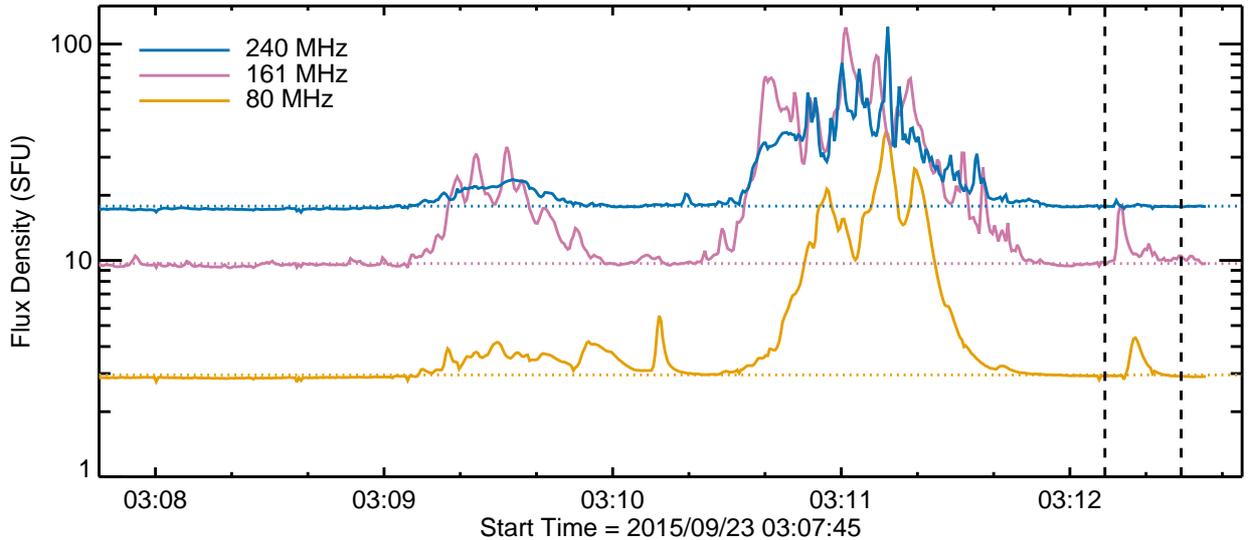}}
\caption{Light curves for the 2015-09-23 observation, shown to illustrate the background level determination. 
Backgrounds (dotted lines) are obtained by taking the median intensity, 
excluding points 2 standard deviations above that, 
and iterating until no more points are excluded. 
The dashed lines mark the burst period from the right column of Fig.~\ref{fig:spectra}.}
\label{fig:baseline}
\end{figure}


\section{Analysis and Results}
\label{analysis}

\subsection{Density Profiles}
\label{density}

Standard plasma emission theory expects type III radiation  
at either the ambient electron plasma frequency ($f_{\rm p}$) or its harmonic 
($2f_{\rm p}$). 
The emission frequency $f$ is related to electron density ($n_e$) in the following way \edit{(in cgs units)}:  

\edit{
\begin{equation} \label{eq:fp}
f = \textrm{N}f_{\rm p} = \textrm{N}\sqrt{\frac{e^2n_e}{\pi{}m_e}} 
~~~\Rightarrow{}~~~ 
n_e = \pi{}m_{e}\left(\frac{f}{\textrm{N}e}\right)^2,
\end{equation} 
}

\noindent where  
$e$ is the electron charge, 
$m_e$ is the electron mass, 
and \edit{N} is either 1 (fundamental) or 2 (harmonic). 
For frequencies in \edit{Hz} and densities in cm\tsp{-3}, $n_e \approx 1.24\e{-8}f^2$ for fundamental 
and $3.10\e{-9}f^2$ for harmonic emission. 

Density can thus be easily extracted given the emission mode and location. 
Unfortunately, neither property is entirely straightforward. 
Harmonic emission is often favored in the corona because being produced above the ambient 
$f_p$ makes it less likely to absorbed \citep{Bastian98} and because type IIIs tend to be 
more weakly circularly polarized than expected for fundamental emission \citep{Dulk80}.
Harmonic emission also implies lower densities by a factor of 4, which are easier to 
reconcile with the large heights often observed (see Section \ref{introduction}). 
However, fundamental--harmonic pairs can be observed near our frequency range (e.g. \citealt{Kontar17}),  
fundamental emission is expected to contribute significantly to interplanetary type III burst spectra (e.g. \citealt{Robinson98}), 
and fundamental emission is thought to be the more efficient process from a theoretical perspective (e.g. \citealt{Li13b,Li14}). 
As described in Section~\ref{introduction}, a source's apparent height may also be 
augmented by \edit{propagation effects}, which we will consider in Section~\ref{propagation}.

We measure source heights at the onset of burst emission, which we define as 
when the total intensity reaches 1.3$\times$ the background level. 
Background levels are determined for each frequency by taking the median intensity, excluding 
points 2 standard deviations above that, and iterating until no more points are excluded. 
Figure~\ref{fig:baseline} shows the result of this baseline procedure for three frequencies from the 2015-09-23 event, 
which is shown because it exhibits the most complicated dynamic spectrum. 
Onset times are represented by circles 
in Figure~\ref{fig:spectra}, and centroids are obtained at these 
times from 2-dimensional (2D) Gaussian fits.
As mentioned in Section \ref{events}, these events were chosen because they appear at the radio limb 
and thus the 2D plane-of-sky positions can reasonably approximate the 
physical altitude. 
Geometrically, these heights are lower limits to the true radial height, but  
propagation effects that increase apparent height 
are likely to be more important than the projection angle (see Section \ref{propagation}).   


\begin{figure}
\centerline{\includegraphics[width=\textwidth,clip=]{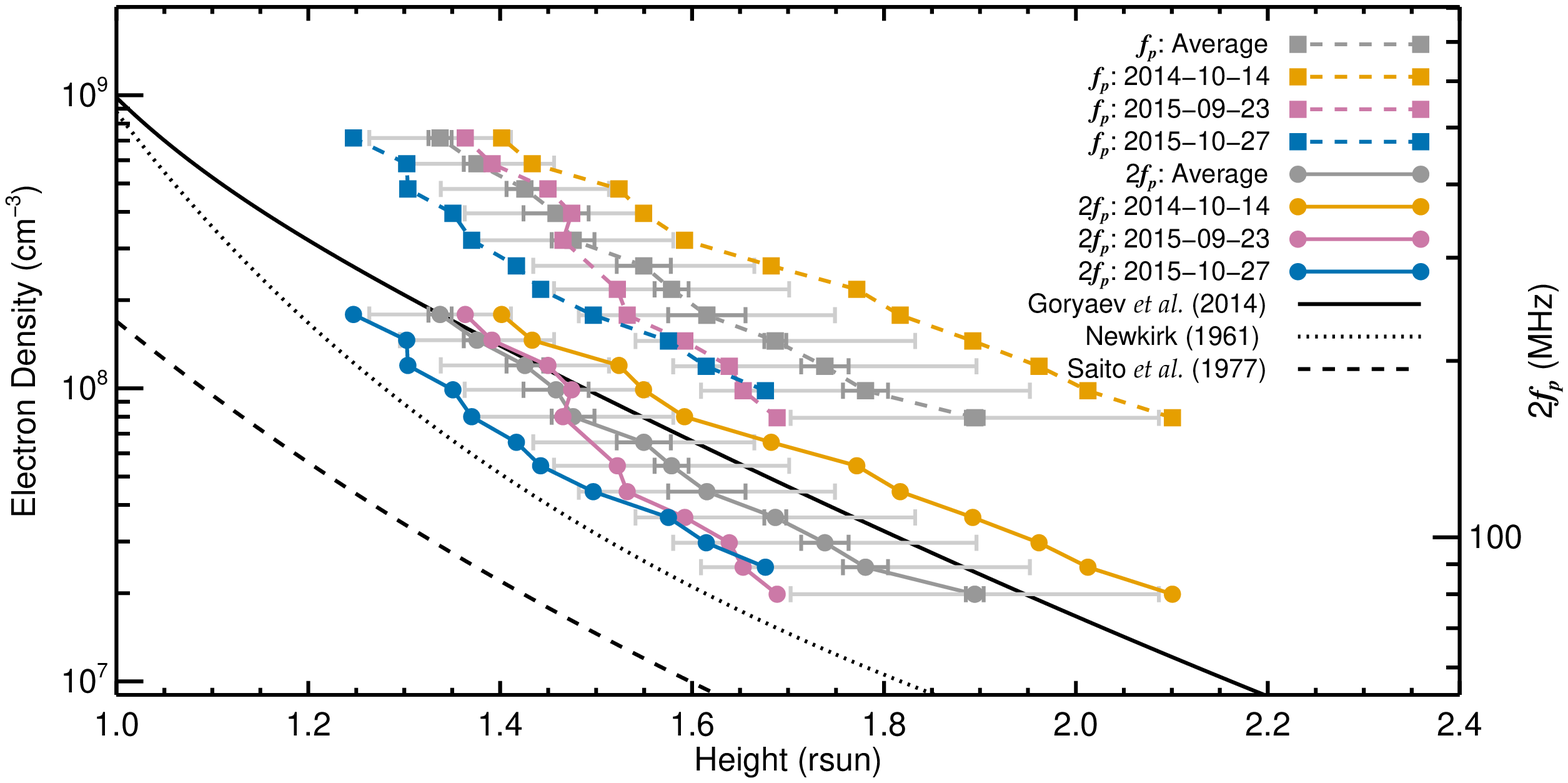}}
\caption{Densities inferred from the type III source positions assuming fundamental ($f_p$; dashed) 
or harmonic ($2f_p$; solid) emission. 
Background coronal models based on white light data near solar minimum \citep{Saito77} and 
maximum \citep{Newkirk61} are shown for comparison, along with a recent streamer model 
based on EUV data \citep{Goryaev14}.
Only the average uncertainties are shown for clarity; the dark gray bars represent the 1$\sigma$ centroid  
variability over the full burst, and the light gray bars represent the major axes of the synthesized beams. 
}
\label{fig:dens1}
\end{figure}  
  

\begin{figure}
\centerline{\includegraphics[width=\textwidth,clip=]{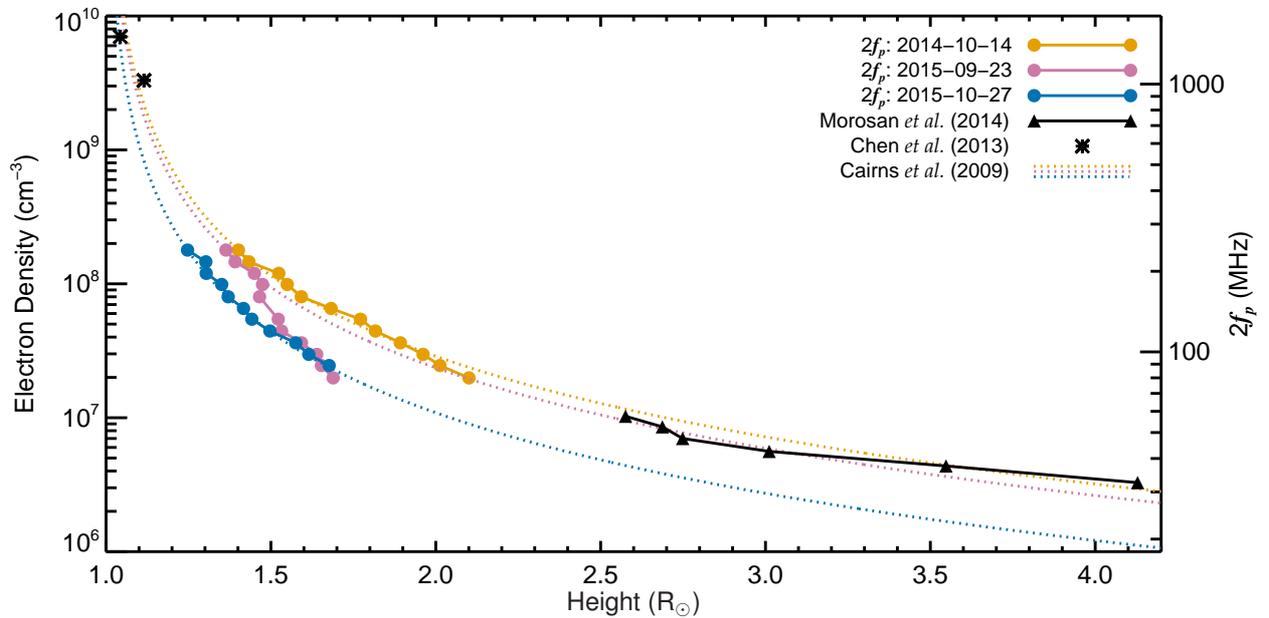}}
\caption{Densities assuming harmonic emission compared to recent type III results at higher \citep{Chen13} 
and lower \citep{Morosan14} frequencies. 
The dotted lines apply the $n_e(r) = C(r - 1)^{-2}$ profile detailed by \citet{Cairns09}, where 
the constant $C$ has been normalized to the density implied by our 240 MHz source positions. 
}
\label{fig:dens2}
\end{figure} 

Figure~\ref{fig:dens1} plots height versus density for both the fundamental and harmonic assumptions. 
Two sets of height uncertainties are shown for the average density profiles. 
The smaller, dark gray error bars reflect the 1$\sigma$ position variability over the full burst durations, and the larger, 
light gray bars reflect the full width at half maximum (FWHM) of the synethesized beam major axes. 
Note that if the source is dominated by a single compact component, which would be a reasonable assumption here, 
then the FWHM resolution uncertainty can be reduced by a factor inversely proportional to the signal-to-noise ratio (SNR) \citep{Lonsdale18,Reid88}. 
\edit{Given our high SNRs, which average 217$\sigma$ at the burst onsets,} this ``spot mapping" approach \edit{typically} results in sub-arcsecond position uncertainties on the apparent source location.
However, spatial shifts may be introduced by changes in the ionosphere between the solar and calibration observation 
times, and more importantly, an apparent source may differ significantly from its actual emission site due to propagation effects (i.e. refraction and scattering). 
For these reasons, we opt to show the more conservative uncertainties outlined above. 

For comparison, Figure~\ref{fig:dens1} includes radial density models from \citet{Saito77}, \citet{Newkirk61}, and \citet{Goryaev14}. 
The \citeauthor{Saito77} profile refers to the equatorial background near solar minimum based on white 
light polarized brightness data, while the \citeauthor{Newkirk61} curve is based on similar data near solar maximum 
and implies the largest densities among ``standard" background models. 
The \citeauthor{Goryaev14} model instead refers to a dense streamer and is based on a 
novel technique using widefield EUV imaging. 
This profile is somewhat elevated above streamer densities 
inferred from contemporary white light (e.g. \citealt{Gibson99}) 
and spectroscopic (e.g. \citealt{Parenti00,Spadaro07}) measurements at similar heights, though 
some earlier white light studies found comparably large streamer densities (e.g. \citealt{Saito67}).
For additional coronal density profiles, see also \citet{Allen47,Koutchmy94,Guhathakurta96,Mann99,Mercier15,Wang17} 
and references therein. 

From Figure~\ref{fig:dens1}, we see that the type III densities assuming fundamental emission are an average  
of 3--4$\times$ higher than the EUV streamer model. 
These values may be unreasonably large, meaning either that the fundamental emission hypothesis is 
not viable here or that fundamental emission \edit{originating} from a lower altitude 
\edit{was observed a larger height due to propagation effects} (see Section \ref{propagation}). 
Assuming harmonic emission, the 2014-10-14 burst implies electron densities of 1.8\e{8} cm\tsp{-3} (240 MHz) 
at \rsolar{1.40} down to 
0.20\e{8} cm\tsp{-3} (80 MHz) at \rsolar{2.10}. 
This represents a moderate ($\sim1.4\times$) enhancement over the \citeauthor{Goryaev14} streamer model 
or a significant ($\sim4.1\times$) enhancement over the \citeauthor{Newkirk61} background. 
The other two events fall between the EUV streamer and solar maximum background models, with the 
2015-10-27 source heights implying densities of 1.8\e{8} cm\tsp{-3} (240 MHz) at \rsolar{1.25} down to 
0.20\e{8} cm\tsp{-3} (80 MHz) at \rsolar{1.68}. 
Note that the 2015-09-23 burst implies an unusually steep density gradient that is not consistent with standard radial 
density models, perhaps because that event 
deviates significantly from the radial direction (see Figure~\ref{fig:overlay}).

Figure~\ref{fig:dens2} shows how our results compare to densities inferred from recent type III imaging at 
higher and lower frequencies, all assuming harmonic emission. 
The high-frequency (1.0--1.5 GHz) results come from \citet{Chen13}, who used the Very Large Array (VLA) 
to find densities around an order of magnitude above the background. 
The low-frequency (30--60 MHz) points were obtained using the 
Low Frequency Array (LOFAR) by \citet{Morosan14}, who also found large enhancements. 
We plot data from their ``Burst 2" (see Figures 3 \& 4) because it began beyond 
our average radio limb height at 80 MHz. 
Their other two events exhibit 60 MHz emission near the optical limb, which may indicate 
that the 2D plane-of-sky positions significantly underestimate the true altitudes 
(i.e. those electron beams may have been inclined toward the observer). 

Figure~\ref{fig:dens2} also includes density curves of the form $n_e(r) = C(r - 1)^{-2}$, where 
$r$ is in solar radii and $C$ is normalized to match the densities implied by our 240 MHz source heights. 
This model was introduced by \citet{Cairns09} based on type III frequency drift rates over 40--180 MHz 
and was subsequently validated over a larger frequency range by \citet{Lobzin10}. 
The \citeauthor{Cairns09} model is somewhat steeper than others over the MWA's height range 
($\sim$ 1.25--2.10\rsolar{}) but becomes more gently-sloping at larger heights,  
effectively bridging the corona to solar wind transition. 
From Figure~\ref{fig:dens2}, we see that this model is a good fit to the 2014-10-14 and 2015-10-27 data. 
The 2015-09-23 event is not well-fit by this or any other standard model, 
which may be attributed to its aforementioned non-radial structure.
Extending these gradients to larger heights matches the LOFAR data fairly well 
and likewise with the VLA data at lower heights, which come from higher frequencies than have been 
examined with this model previously. 

\subsection{Electron Beam Kinematics}
\label{speed}


\begin{figure}
\centerline{\includegraphics[width=\textwidth,clip=]{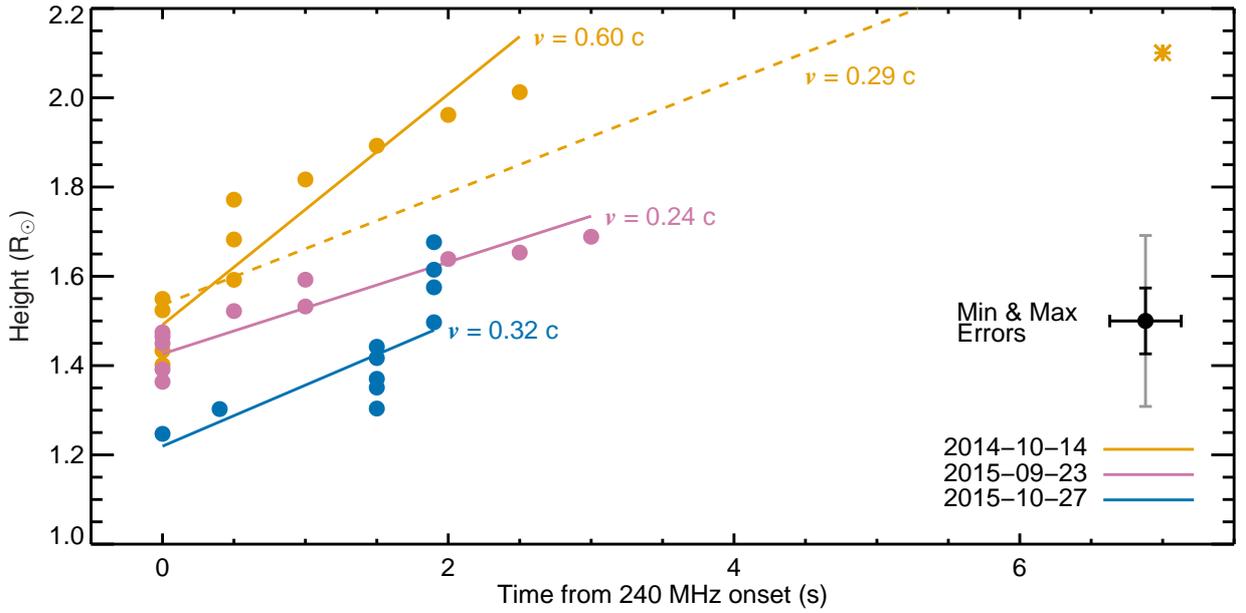}}
\caption{Exciter speed estimates from the time- and frequency-varying source positions.
The dashed orange line includes the high time outlier (orange asterisk). 
The uncertainties shown in the lower right are the same for a given frequency and reflect the time and spatial resolutions. 
The black bar represents the smallest synthesized beam size at 240 MHz (corresponding to the lower-left points), 
and the gray bar represents the largest beam size at 80 MHz (corresponding to the upper-right points). 
}
\label{fig:speeds}
\end{figure}  

Type III beams speeds are known primarily from frequency drift rates ($df/dt$) observed in dynamic spectra. 
Assuming either fundamental or harmonic emission, a given burst frequency can be straightforwardly 
converted into a radial height given a density model $n_e(r)$, and $df/dt$ then becomes $dr/dt$. 
The literature includes a wide range of values using this technique, reflecting the variability  
among models as well as any intrinsic variability in electron speed. 
Modest fractions of light speed are typically inferred from drift rates of coronal bursts ($\sim$0.1--0.4 c; 
\citealt{Alvarez73,Aschwanden95,Melendez99,Kishore17}), though speeds larger 
than 0.5 c have been reported by some studies \citep{Poquerusse94,Klassen03,Carley16}. 
Our imaging observations allow us to measure the exciter speed without assuming $n_e(r)$ by following the 
apparent height progression of type III sources at different frequencies. 

As in the previous section, we obtain radial heights from centroid positions at the 
onset of burst emission for each frequency. 
These data are plotted in Figure~\ref{fig:speeds} along with linear least-squares fits to the speed using   
the time and spatial resolutions as uncertainties. 
The 2014-10-14 event exhibits an anomalously \edit{late} onset time at 80 MHz 
\edit{(see the circles in Figure~\ref{fig:spectra}a and the orange asterisk in Figure~\ref{fig:speeds}). 
This is likely due to the diminished intensity at that frequency, which precludes an appropriate comparison 
to the onset times at higher frequencies where the burst is much more intense. 
Figure~\ref{fig:speeds} shows fits both including (0.29 c) and excluding (0.60 c) the 80 MHz point for 
the 2014-10-14 event, and the latter value is used in the discussion to follow because of the better overall fit. 
Note that while the onset of 80 MHz emission is at a later time than expected given the prior frequency progression, 
the source location is consistent with the other channels and thus 
its inclusion not does not impact the inferred density profile from Figures~\ref{fig:dens1} and \ref{fig:dens2}.}

We find an average speed across events of 0.39 c, which is consistent with results from other imaging observations.
The same strategy was recently employed at lower frequencies by \citet{Morosan14}, who found an 
average of 0.45 c. 
\citet{McCauley17} indirectly inferred a beam speed of 0.2 c from MWA imaging. 
\citet{Chen13} also tracked centroid positions at higher frequencies, though 
in projection across the disk, finding 0.3 c. 
\edit{\citet{Mann18} recently examined the apparent speeds of three temporally adjacent type III bursts imaged by 
LOFAR. They find that the sources do not propagate with uniform speed, with each burst exhibiting an  
acceleration in apparent height, and they conclude that the exciting electron beams must have broad velocity distributions. 
From Figure~\ref{fig:speeds}, we observe an apparent acceleration only for one event (2015-10-27), with the other two events 
exhibiting the opposite trend to some extent. 
However, our data are consistent with \citet{Mann18} in that a uniform speed is not a particularly good fit for any of our events, but 
the MWA's 0.5 s temporal resolution limits our ability to characterize the source speeds in great detail.}


\begin{table}
\caption{Imaging Beam Speeds vs $df/dt$ Model Predictions}
\label{tab:speeds}
\begin{tabular}{l|c|ccc|c}
\hline
& \multicolumn{5}{c}{Beam Speed (c)} \\
&  & \multicolumn{3}{c|}{Assuming $f_p$ -- $2f_p$ emission} &  \\
Event & Imaging & Goryaev \textit{et al.} & Newkirk & Saito \textit{et al.}\tabnote{$f_p$ case not viable because model does not include densities above $f_p \approx$ 116 MHz.} & Cairns \textit{et al.}\tabnote{Model normalized to match the densities implied by our 240 MHz heights.} \\
\hline
2014-10-14\tabnote{Excludes the 80 MHz outlier (orange asterisk in Fig. 10).} & 0.60 $\pm$ 0.13 & 0.38 -- 0.45 & 0.22 -- 0.31 & *** -- 0.30 & 0.58 \\
2015-09-23 & 0.24 $\pm$ 0.10 & 0.34 -- 0.40 & 0.20 -- 0.28 & *** -- 0.27 & 0.50 \\
2015-10-27 & 0.32 $\pm$ 0.12 & 0.44 -- 0.55 & 0.26 -- 0.36 & *** -- 0.48 & 0.40 \\
\hline
\end{tabular}
\end{table}

Taken together, we see that speeds measured from imaging observations tend to produce values 
at the higher end of what is typical for $df/dt$ inferences. 
We compare the two approaches for the same events in Table~\ref{tab:speeds} using the same 
models shown in Figure~\ref{fig:dens1}. 
We also include speeds derived using the \citet{Cairns09} model, normalized to the densities 
implied by our 240 MHz source heights.  
These values are separated from the others in Table~\ref{tab:speeds} because the normalization 
precludes direct comparisons to the other models. 
The $df/dt$-inferred speeds are consistently smaller than the imaging results for the 2014-10-14 event, which 
was also true for the bursts studied by \citet{Morosan14}, but there is no major difference between the 
two approaches for our other events given the range of values. 
Note that this comparison is arguably a less direct version of the  
height versus density comparison from the previous section in that 
the extent to which the imaging and model-dependent $df/dt$ speeds agree unsurprisingly mirrors the extent 
to which the density profiles themselves agree.  
The 2014-10-14 speeds are closest to those derived using $n_e(r)$ from \citeauthor{Goryaev14}, and 
the 2015-10-27 result is closest to the \citeauthor{Newkirk61}-derived speed, both assuming 
harmonic emission, because those density profiles are most closely matched in Figure~\ref{fig:dens1}.  
Likewise, the speeds from those events agree well with $df/dt$ speeds obtained using the 
normalized \citeauthor{Cairns09} curves because a $C(r - 1)^{-2}$ gradient fits those data nicely. 
The 2015-09-23 speed is between the two values derived using the \citeauthor{Newkirk61} model assuming either 
fundamental and harmonic emission, but this may be coincidence given that the modeled and observed 
density profiles are widely discrepant. 
That event's non-radial 
profile may also prevent meaningful agreement with any simple $n_e(r)$ model (see Figure~\ref{fig:overlay}). 


\subsection{Propagation Effects}
\label{propagation}

As described in Section \ref{introduction}, a number of authors have argued that radio propagation 
effects, namely refraction and scattering, can explain the large source heights frequently exhibited by type III bursts. 
\citet{Bougeret77} introduced the idea of scattering by overdense fibers in the 
context of radio burst morphologies, and \citet{Stewart74} suggested that type III 
emission may be produced in underdense flux tubes as a way of explaining observed 
harmonic--fundamental ratios. 
These two concepts were combined by \citet{Duncan79}, who introduced the 
term \textit{ducting} to refer to radiation that is produced in an underdense environment and subsequently 
guided to a larger height by reflections against a surrounding ``wall" of much 
higher-density material, which eventually becomes transparent with sufficient altitude.   
\edit{While plausible, this concept generalizes poorly in that electron beams are not 
expected to be found preferentially within coherent sets of low-density structures that would be conducive to ducting.}

\edit{\citet{Robinson83} addressed this by showing}
that random reflections against overdense fibers can 
have the same effect of elevating an observed burst site above 
its true origin, but without requiring any peculiarities of the emission site (i.e. low-density). 
\edit{Because the high density fibers known to permeate the corona are not randomly 
arranged and are generally radial, random scattering against them does not randomly modulate 
the aggregate ray path--scattering instead tends to guide the emission outward to larger heights in a manner 
that is analogous to the classic ducting scenario.
For this reason, other authors (e.g. \citealt{Poquerusse88}) have chosen to retain \textit{ducting} 
to refer to the similar but more general impact of scattering, without implying that the emission is 
guided within a particular density structure as originally proposed by \citet{Duncan79}. 
Here, we will simply refer to \textit{scattering} to avoid potential confusion between the two concepts.} 

\edit{After being scattered for the last time upon reaching a height with sufficiently-low densities, a radio wave will 
then be refracted through the corona before reaching an observer, further shifting the source location. 
As the coronal density gradient generally decreases radially, radio waves will tend to refract toward 
to the radial direction such that a source originating at the limb will appear at a somewhat lower height than its origin, which  
could be either the actual emission site (e.g. \citealt{Stewart76}) or, more likely, the point of last scatter (e.g. \citealt{Mann18}). 
Accounting for the refractive shift, which becomes larger with decreasing frequency, therefore requires that the emission be generated at 
or scattered to an even larger height than is implied by the observed source location. 
Recent results on this topic from \citet{Mann18} will be discussed in the next section.} 
 
Propagation effects are also thought to be important to the observed structure of the quiescent corona, 
where the dominant emission mechanism is thermal 
bremsstrahlung (free-free) radiation at MWA frequencies.   
Outside of coronal holes, this emission is expected to be in or close to the optically thick regime (e.g. \citealt{Kundu82,Gibson16}), which means that the observed brightness temperature 
should be the same as the coronal temperature.
However, well-calibrated 2D measurements have generally found lower brightness 
temperatures than expected from temperatures derived at other wavelengths (see review by \citealt{Lantos99}). 
Additionally, the size of the corona appears to be larger than expected at low frequencies 
(e.g. \citealt{Aubier71,Thejappa92,Sastry94,Subramanian04,Ramesh06}). 
The prevailing explanation for these effects is also scattering by density inhomogeneities 
(e.g. \citealt{Melrose88,Alissandrakis94,Thejappa08})\edit{, though the refractive effect described in the 
previous paragraph is also important \citep{Thejappa08}.}

Thus, the \edit{scattering} process that may act to elevate type III sources also 
affects quiescent emission, increasing the apparent size of the corona.
We will take advantage of this by using the difference in extent between 
observed and modeled quiescent emission 
as a proxy for the net effect of propagation effects on our type III source heights. 
\edit{This approach is limited in that, although both are related to scattering, the extent 
to which the magnitudes of these two phenomena are related is unclear. 
In particular, previous studies on the broadening of the radio Sun by scattering 
have considered random density inhomogeneities as opposed to the more realistic 
case of high density fibers capable of producing the ducting-like effect for type III sources.}
 

\begin{figure}
\centerline{\includegraphics[width=\textwidth,clip=]{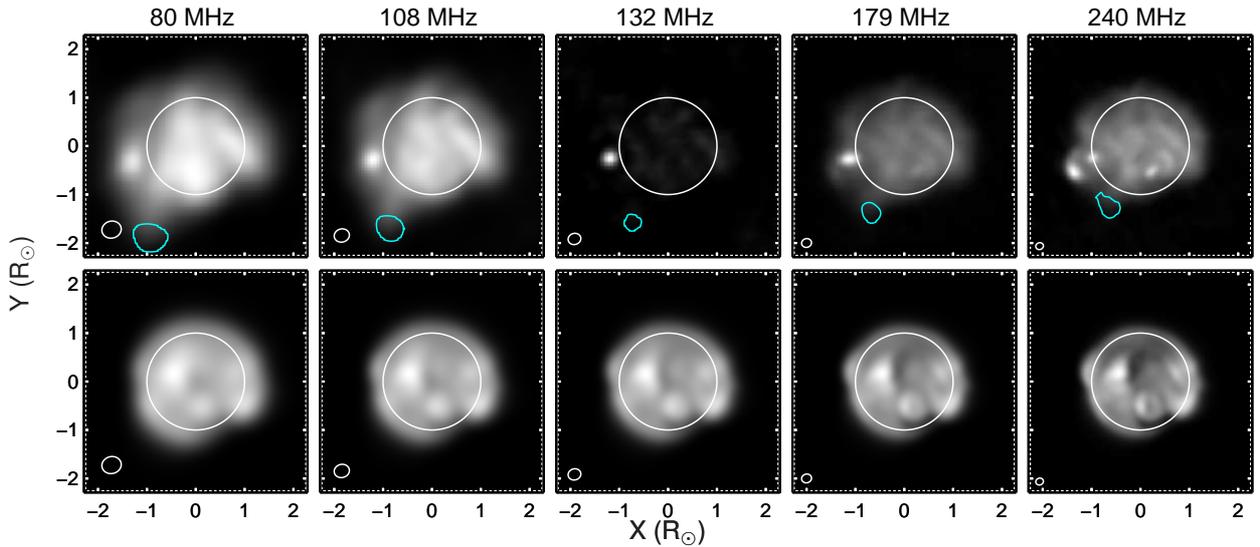}}
\caption{MWA background images for the 2014-10-14 event (top) and corresponding MAS--FORWARD synthetic 
images convolved with the MWA beam (bottom). 
Beam ellipses are shown in the lower-left corners, 
and the cyan curves are the 50\% burst contours from Figs.~\ref{fig:centroids} \& \ref{fig:overlay}. 
This day is shown because thermal emission is only barely distinguishable at 132 MHz, precluding 
the Fig.~\ref{fig:offset} analysis at that frequency, which was not the case for any other event-channel combination.
}
\label{fig:forward}
\end{figure}  


\begin{figure}
\centerline{\includegraphics[width=\textwidth,clip=]{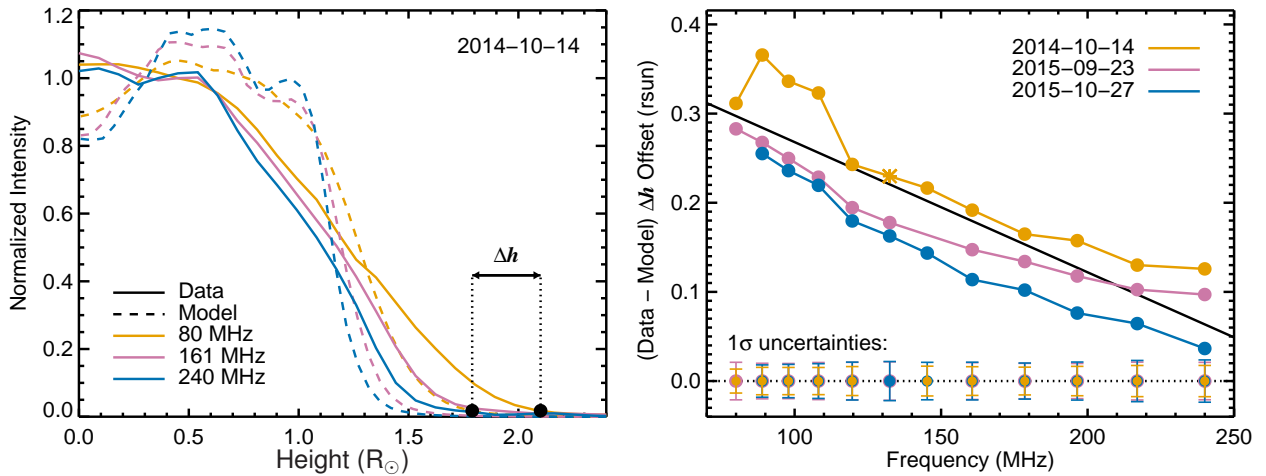}}
\caption{\textit{Left}: Average intensity versus radial distance obtained from the Fig.~\ref{fig:forward} images 
and normalized by the median value below \rsolar{1}.  $\Delta{h}$ refers to the height offset between the observed and 
modeled intensity profiles at the apparent type III burst height at 80 MHz.  
\textit{Right}: $\Delta{h}$ for each frequency and event. 
An orange asterisk marks the one instance where data was available but a measurement could not be made 
because the thermal background emission was not well-detected (see Fig.~\ref{fig:forward}), so an average of the 
adjacent points is used. 
The uncertainties reflect the sensitivity of $\Delta{h}$ to the normalization choice in the left panel (see Sec.~\ref{propagation}).  
}
\label{fig:offset}
\end{figure}   

Figure~\ref{fig:forward} shows the observed background emission versus synthetic images  
based on a global MHD model. 
The MWA images are obtained by averaging every frame with a total intensity less than 
2$\sigma$ above the background level, which is determined via the procedure shown in Figure~\ref{fig:baseline}.
Synthetic images are obtained using
FORWARD \citep{Gibson16}, a 
software suite that calculates the expected bremsstrahlung and gyroresonance emission 
given a model atmosphere. 
We use the Magnetohydrodynamic Algorithm outside a Sphere 
(MAS; \citealt{Lionello09}) medium resolution (\texttt{hmi\_mast\_mas\_std\_0201}) model, which extrapolates 
the coronal magnetic field from photospheric magnetograms (e.g. \citealt{Miki99}) 
and applies a heating model adapted from \citet{Schrijver04} to compute density and temperature. 

\citet{McCauley17} established the use of these model images for flux calibration and included a qualitative 
comparison to MWA observations. 
As in the aforementioned literature, 
the radial extent of the corona is somewhat larger in the observations than 
in the beam-convolved model images. 
To quantify this difference, we divide both image sets into concentric rings about Sun-center. 
The average intensity within each ring is plotted against its radial distance in 
the left panel of Figure~\ref{fig:offset}, where the intensities have been normalized by the 
median value below \rsolar{1}. 
We then measure the offset $\Delta{h}$ between the observed and modeled profiles at the 
heights obtained from the type III positions. 
$\Delta{h}$ is sensitive to how the intensity curves are normalized, and we quantify 
this uncertainty by repeating the procedure for 10 different normalization factors that 
reflect the median intensities within radial bins of width \rsolar{0.1} from 0 to \rsolar{1}.
The right panel of Figure~\ref{fig:offset} plots the $\Delta{h}$ results for each event,   
which have 1$\sigma$ uncertainties of less than $\pm$\rsolar{0.025}.
The offset appears to depend roughly  
linearly on frequency, with larger offsets at lower frequencies. 
Fitting a line through all of the points, we find that: 
\begin{equation}
\label{eq:dh}
\Delta{h} \approx{} -1.5\e{-3}f + 0.41; ~80\leq{} f \leq{240}~\rm{MHz}
\end{equation}

\noindent where $\Delta{h}$ is in solar radii and $f$ is in MHz. 
This yields \rsolar{0.30} at 80 MHz and \rsolar{0.06} at 240 MHz. 
We do not expect this expression to be relevant much outside of the prescribed frequency range, 
but extrapolating slightly, we obtain \rsolar{0.32} at 60 MHz. This value is a bit more than 
half of the $<$ \rsolar{0.56} limit found by \citet{Poquerusse88}. 

\citeauthor{Poquerusse88}, and others who have quantified the scattering effect (e.g. \citealt{Robinson83}), 
obtained their results by computing ray trajectories through a model corona. 
That approach allows a fuller understanding of the propagation physics, 
but the result is dependent on the 
assumed concentration and distribution of high density fibers, which are not well constrained. 
Our critical assumption is that emission produced at significantly lower heights would be absorbed, as 
would be the case in our optically-thick model corona. 
However, low coronal brightness temperatures could also be 
explained by lower opacities (e.g. \citealt{Mercier09}) or a low filling factor, which would allow burst emission 
to escape from lower heights and lead us to underestimate the potential impact of \edit{propagation effects}. 

\begin{figure}
\centerline{\includegraphics[width=\textwidth,clip=]{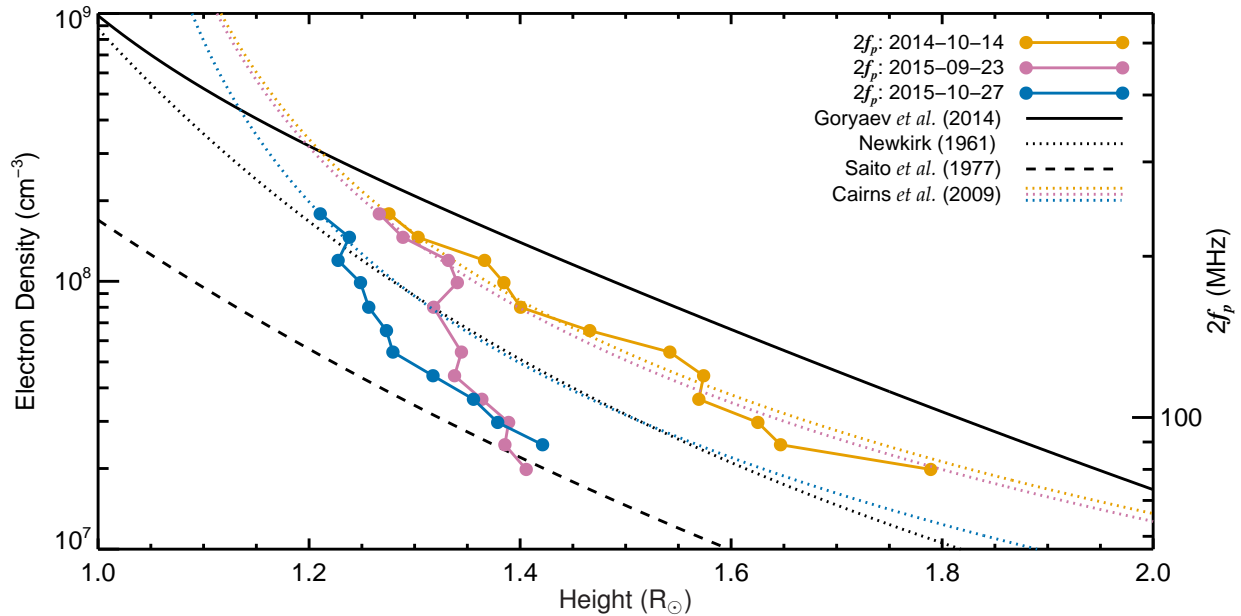}}
\caption{Imaging density profiles after applying the $\Delta{h}$ offsets from Fig.~\ref{fig:offset} and 
assuming harmonic emission. 
Model curves are as in Figs.~\ref{fig:dens1} \& \ref{fig:dens2}. 
}
\label{fig:dens3}
\end{figure}

Figure~\ref{fig:dens3} shows how the Figure~\ref{fig:dens1} density results change after subtracting the 
height offsets from Figure~\ref{fig:offset}. 
The 2014-10-14 harmonic ($2f_p$) profile remains reasonable with the offsets, lying just below the 
\citet{Goryaev14} model instead of just above it, while the 
fundamental emission densities for that event would still be quite large. 
Given that the \citeauthor{Goryaev14} model is among the highest-density streamer models 
in the literature, we conclude that harmonic emission from a beam traveling 
along an overdense structure is consistent with 2014-10-14 data. 
Our assessment for this event is that  
propagation effects may contribute to some but not all the apparent density enhancement. 

The other two events exhibit unusually steep density profiles once the offsets are subtracted.  
That was true also for the original 2015-09-23 results, which we attributed to its non-radial 
trajectory in Section~\ref{density}. 
However, the original 2015-10-27 densities were well-matched to the 
\citet{Cairns09} $C(r - 1)^{-2}$ gradient but become too steep to match any standard density gradient
with the inferred offsets. 
This may simply reflect the intrinsic density gradient of the particular structure.
Alternatively, it is possible that we have over- or underestimated the impact of \edit{propagation effects} at the low or high end of our 
frequency range, respectively.  
However, the frequency dependence of scattering \edit{and refraction} means that any treatment will  
steepen the density gradient. 
Aside from their slopes, the offsets bring the densities implied by both bursts to generally within the 
normal background range assuming harmonic emission or to a moderately enhanced level 
assuming fundamental emission.  


\section{Discussion} %
\label{discussion} %

The previous section suggests that propagation effects can partially explain 
the apparent density enhancements implied by our type III source heights. 
Assuming harmonic emission, our estimates for the potential magnitude of \edit{propagation effects} bring 
the densities to within normal background levels for two events, while one 
event remains enhanced at a level consistent with a dense streamer. 
In Section~\ref{density}, we showed that our original density profiles are 
consistent with those found from recent type III burst studies at lower  
and higher frequencies, which together are well-fit by the 
\citet{Cairns09} $C(r - 1)^{-2}$ gradient. 
Both the low- and high-frequency studies conclude that their large densities 
imply electron beams traveling along overdense structures, 
but neither consider the impact of propagation effects.  

\citet{Morosan14} propose a variation of the overdense hypothesis based 
on their 30--60 MHz LOFAR observations. They suggest that 
the passage of a CME just prior to an electron beam's arrival may compress streamer plasma 
enough to facilitate type III emission at unusually large heights. 
This was consistent with their events being associated with a CME and could be relevant to 
our 2015-09-23 event, \edit{which was also preceded by a CME} (see~Section \ref{context}). 
While this interpretation is plausible, 
we do not think such special conditions need to be invoked given that the densities inferred from the 
\citeauthor{Morosan14} results are consistent with ours (Figure~\ref{fig:dens2}) and are broadly consistent with the large  
type III source heights found using the previous generations of low frequency instruments (e.g. \citealt{Wild59,Morimoto64,Malitson66,Stewart76,Kundu84}). 
\edit{Propagation effects} seem particularly likely to have contributed (at least partially) to their inferred density enhancement, as the 
effects become stronger with decreasing frequency. 

\edit{Recently, \citet{Mann18} also examined the heights of type III sources observed at the limb by LOFAR. 
After accounting for the refractive effect described in Section~\ref{propagation}, and relying on scattering to direct 
emission toward the observer at large heights prior to being refracted, 
their results imply a density enhancement of around 3.3$\times$ over the \citet{Newkirk61} density 
model, assuming fundamental emission. 
Incorporating our offsets from Section~\ref{propagation} gives us an average enhancement of 4.6$\times$ 
over the same model across our three events, also assuming fundamental emission.
Our results are therefore consistent with those of \citet{Mann18}, though our attempts quantify the impact
of propagation effects are quite different.}

\citet{Chen13} also found large densities using VLA data at higher frequencies (1.0--1.5 GHz).
Scattering is also thought to be important at high frequencies given the apparent lack 
of small-scale structure \citep{Bastian94}, but the 
extent to which \edit{scattering may also elevate radio sources} in that regime has 
not been addressed to our knowledge. 
\citeauthor{Chen13} observed an on-disk event, from which they obtain source heights by comparing 
their projected positions to stereoscopic observations of an associated EUV jet, which is assumed 
to have the same inclination as the type III electron beam. 
This method would also be impacted by any source shifting caused by \edit{scattering}. 
Although these shifts would be much smaller than at low frequencies, the background 
gradient is much steeper, so a reasonably small shift may still strongly influence the 
inferred density relative to the background. 

If we accept the densities obtained at higher frequencies, albeit from just one example, 
then their consistency with low frequency observations in general is striking. 
As we describe in Section~\ref{introduction}, the community largely came to favor propagation effects 
over the overdense hypothesis in the 1980s, and 
the topic has not had much consideration since. 
If new observations at low heights (high frequencies) also suggest beams moving 
preferentially along dense structures, then it begs the question of whether or not that interpretation is again  
viable at larger heights (lower frequencies). 
In that case, this would need to be reconciled with the 
fact that electron beams have not been found to be preferentially associated 
with particularly high density regions in \textit{in situ} solar wind measurements \citep{Steinberg84}, along with 
the evidence for other impacts of scattering such as angular broadening 
(e.g. \citealt{Steinberg85,Bastian94,Ingale15}).

Selection effects may be relevant, as radiation produced well above the ambient plasma frequency is 
less likely to be absorbed before reaching the observer. 
Thus, coronal type III bursts may imply high densities because beams traveling along dense structures are 
likelier to be observed. 
Type III bursts also have a range of starting frequencies, 
which has been interpreted in terms of a range in acceleration (i.e. reconnection) heights that 
are often larger than those inferred from X-ray observations \citep{Reid14b}.
Alternatively, a beam may be accelerated at a smaller height than is implied by 
the resultant type III starting frequency due to 
unfavorable radiation escape conditions (absorption) below the apparent starting height. 
Simulations also suggest that electron beams may travel a significant distance before producing 
observable emission \citep{Li13b,Li14} and that they may be radio loud at 
some frequencies but not at others due to variations in the ambient 
density \citep{Li12,Loi14} and/or temperature \citep{Li11b,Li11}.

The magnetic structures along which electron beams travel also evolve with distance from 
the Sun. 
A popular open flux tube model is an expanding funnel that is thin at the base of the corona 
and increasingly less so into the solar wind (e.g. \citealt{Byhring08,He08,Pucci10}). 
Such structures may allow a dense flux tube to become less dense relative 
to the background as it expands with height. 
Moreover, a beam following a particular magnetic field line from the corona into the high-$\beta$ solar wind may not 
necessarily encounter a coherent density structure throughout the heliosphere. 
Turbulent mixing, corotation interaction regions, CMEs, and other effects influence 
solar wind density such that it is not obvious that an electron beam traversing an overdense structure near the Sun 
should also be moving in an overdense structure at large heliocentric distances. 

We also note that one of the main conclusions from many of the type III studies referenced here is unchanged 
in either the overdense or \edit{propagation effects} scenarios. 
Both cases require a very fibrous corona that can supply dense structures 
along which beams may travel and/or dense structures capable of \edit{scattering} radio emission \edit{to larger heights}.  
This consistent with the fine structure known from eclipse observations (e.g. \citealt{Woo07}) that has more recently 
been evidenced by EUV observations. 
For instance, analyses of a sungrazing comet \citep{Raymond14} and of EUV spectra \citep{Hahn16} 
independently suggest large density contrasts ($\gtrsim$ 3--10) between neighboring flux tubes in 
regions where the structures themselves are undetected. 
As our understanding of such fine structure improves, better constraints can be placed on them 
for the purpose of modeling the impact of \edit{propagation effects} on radio sources. 

 
\section{Conclusion} %
\label{conclusion} %

We presented imaging of three isolated type III bursts observed at the 
limb on different days using the MWA.
Each event is associated with a white light streamer 
and plausibly associated with EUV rays that exhibit activity around the 
time of the radio bursts. 
Assuming harmonic plasma emission, density profiles derived from 
the source heights imply enhancements of 
$\sim$2.4--5.4$\times$ over background levels. 
This corresponds to electron densities 
of 1.8\e{8} cm\tsp{-3} (240 MHz) 
down to 0.20\e{8} cm\tsp{-3} (80 MHz) at average heights of 1.3 to \rsolar{1.9}.
These values are consistent with the highest streamer densities inferred from other 
wavelengths and with the large radio source heights found using older instruments. 
The densities are also consistent with recent type III results at higher and lower frequencies, 
which combined are well-fit by a $C(r - 1)^{-2}$ gradient. 
By comparing the extent of the radio limb to model predictions, we estimated that 
\edit{radio propagation effects, principally the ducting-like effect of random scattering by high density fibers,} 
may be responsible for 0.06--0.30\rsolar{} 
of our apparent source heights. 
This shift brings the results from 2 of our 3 events to within a standard range of background densities. 
We therefore conclude that propagation effects can partially explain the apparent density enhancements but 
that beams moving along overdense structures cannot be ruled out. 
We also used the imaging data to estimate electron beam speeds of 0.24--0.60 c. 

\begin{acks}  
Support for this work was provided by the Australian Government through an Endeavour Postgraduate Scholarship. 
We thank Stephen White and Don Melrose for helpful discussions 
\edit{and the anonymous referee for their constructive comments}.
This scientific work makes use of the Murchison Radio-astronomy Observatory (MRO), operated by 
the Commonwealth Scientific and Industrial Research Organisation (CSIRO). 
We acknowledge the Wajarri Yamatji people as the traditional owners of the Observatory site. 
Support for the operation of the MWA is provided by the Australian Government's  
National Collaborative Research Infrastructure Strategy (NCRIS), 
under a contract to Curtin University administered by Astronomy Australia Limited. 
We acknowledge the Pawsey Supercomputing Centre, which is supported by the 
Western Australian and Australian Governments.
The SDO is a National Aeronautics and Space Administration (NASA) satellite, and 
we acknowledge the AIA science team for providing open 
access to data and software. 
The SOHO/LASCO data used here are produced by a consortium of the Naval Research Laboratory (USA), 
Max-Planck-Institut fuer Aeronomie (Germany), Laboratoire d'Astronomie (France), 
and the University of Birmingham (UK). 
SOHO is a project of international cooperation between ESA and NASA.
This research has also made use of NASA's Astrophysics Data System (ADS), 
along with JHelioviewer \citep{Muller17} and the Virtual Solar Observatory (VSO, \citealt{Hill09}). 
\end{acks}


  
\bibliographystyle{spr-mp-sola}

\tracingmacros=2
\bibliography{mwa_type3_densities_references}  

\IfFileExists{\jobname.bbl}{} {\typeout{}
\typeout{****************************************************}
\typeout{****************************************************}
\typeout{** Please run "bibtex \jobname" to obtain} \typeout{**
the bibliography and then re-run LaTeX} \typeout{** twice to fix
the references !}
\typeout{****************************************************}
\typeout{****************************************************}
\typeout{}}

\end{article} 

\end{document}